\documentclass[11pt]{article}
\usepackage[colorlinks=false, linkcolor=black, citecolor=black]{hyperref}
\usepackage{graphicx}
\usepackage{subfig}
\usepackage{amstext,amsmath,amssymb,amsfonts,bbm,amsthm}
\usepackage{authblk}

\def\sec#1{{Section~\ref{#1}}}
\def\eqs#1{{Eq.~(\ref{#1})}}

\def\fig#1{{Fig.~\ref{#1}}}

\def\be{\begin{equation}}
\def\la{\label}
\def\ee{\end{equation}}
\def\bes{\begin{eqnarray}}
\def\ees{\end{eqnarray}}
\def\bln{\begin{align}}
\def\eln{\end{align}}
\def\bwt{\begin{widetext}}
\def\ewt{\end{widetext}}

\def\nn{\nonumber}

\title{Coulomb screening in linear coasting nucleosynthesis}

\author[1]{Parminder Singh\thanks{parminderofficial.du@gmail.com}}
\author[2]{Daksh Lohiya\thanks{dl116@cam.ac.uk}}
\affil[1]{Dronacharya Government College, Gurugram University, Gurugram, INDIA}
\affil[2]{Department of Physics and Astrophysics, University of Delhi,\\ Delhi 110007, INDIA.}

\date{\today}

\begin{document}

\maketitle

\begin{abstract}
We investigate the impact of coulomb screening on primordial nucleosynthesis
in a universe having scale factor that evolves linearly with time. Coulomb screening 
affects primordial nucleosynthesis via enhancement of thermonuclear reaction rates. This enhancement is 
determined by the solving Poisson equation within the context of mean field theory
(under appropriate conditions during the primordial
nucleosynthesis). Using these results, we claim that the mean field 
estimates of coulomb screening hardly affect the 
predicted element abundances and nucleosynthesis parameters$, \{\eta_9,\xi_e\}$. 
The deviations from mean field estimates are also studied in detail by
boosting genuine screening results with the screening parameter ($\omega_s$).
These deviations show negligible effect on the element abundances 
and on nucleosynthesis parameters. This work thus rules out the coulomb screening effects on primordial
nucleosynthesis in slow evolving models and confirms that constraints in ref. \cite{parm2014} on nucleosynthesis parameters remain unaltered. 
\end{abstract}

\section{Introduction}
\label{intro}

``Power Law Cosmologies" in which the scale factor evolves as $a(t)\sim t^n$ with $n\geq1$ have emerged as a potential alternative to the standard cosmological models. It is being free of issues found in the standard scenario like the flatness problem, horizon problem, identification of dark matter components etc. \cite{man1,man2}. The power law models can be motivated either in the framework of conformal gravity \cite{man2} or by considering non-minimally coupled unstable scalar fields in the FRW scenario \cite{ford}. Particularly interesting model in this class of cosmologies is the linearly coasting universe with $n = 1$. Even though quite simple, this special case has been found to be concordant with the many observational features such as the SNe1a data \cite{adev01,gset1}, lensing statistics \cite{adev02} and the primordial nucleosynthesis \cite{parm2014,lohiya,chardin}.

It was shown in ref.\cite{parm2014} that primordial nucleosynthesis in a Linearly Coasting Cosmology (LCC) can \emph{successfully} achieve the observed levels of $^4$He \cite{izotov1} and the minimum metallicity levels (Z$_{cr}\sim 10^{-6}Z_\odot$) within the allowed ranges of baryonic density ($\Omega_B$) and puts a constraint on the electron neutrino degeneracy parameter ($\xi_e$). Such constraints on $\Omega_B$ saturates the matter content of universe by baryons alone and eliminates ``Dark Matter Problem". The order of primordial metallicity achieved in the process also removes the need of hypothetical PopIII stars of standard cosmological models which are required for the fragmentation, cooling and  formation process of proto-stellar clouds as they form observed lowest metal stars \cite{sch2006,omu2005,escan,jmesc}.

It has been argued in ref. \cite{pranavscr} that coulomb screening during the primordial nucleosynthesis may have an effect on the nuclear reaction rates and can enhance the nucleosynthesis process which could change the estimates of the parameters $\Omega_B$ in LCC.
Such a screening mechanism can be studied as follows. At the epoch of nucleosynthesis, the temperature of the universe is of the order of MeVs and the universe consists of a hot plasma containing photons, electron-positron pairs, neutrinos and antineutrinos of three flavours and a very small amount of baryons \cite{parm2014}. In this plasma, charged particle number density is dominated by the e$^{\pm}$ pairs \cite{pranavscr}. Any two nuclides having charges $Z_1 e$ and  $Z_2 e$ will be screened by the negatively charged electrons of the plasma. This screening of nuclei charges by the electron cloud reduces the coulomb potential of these charges and would enhance the rate of nuclei interaction.
This has been studied in great detail for stellar interiors in refs. \cite{salpeter,salpeter1,itoh1,itoh2,schman1,schman2,keller}.
One can also calculate the coulomb screening for the early universe by using the well-known mean field approximation  
but with appropriate modifications.
This can then be incorporated in the nuclear reaction rates.
This is a completely different task than the standard BBN where the reaction network contains only 88 reactions and 26 nuclide \cite{wag69,wag73,kaw88,kaw92,itoh,wang}
because only light elements can be synthesized up to the observed levels due to rapid expansion of universe 
and a very little amount of metallicity is produced (Z$\sim$ $\cal{O}$(10$^{-16}$) \cite{iocco}.
The modified nucleosynthesis code for the linear evolution of the scale factor \cite{parm} on the other 
hand considers a heavy network of 557 reaction rates and 130 nuclide since heavy elements (eg. CNO)
can easily be produced upto appreciable levels in this framework. This has its roots in the  
slower expansion rates of the universe \cite{parm2014}. In order to determine the
correct amount of nuclide abundance,
this nucleosynthesis network requires details of nuclear reaction rates $\langle\sigma v\rangle$,  which depend upon cross-section 
`$\sigma$' and relative velocity `$v$' of interacting nuclides governed by the Maxwell-Boltzmann distribution for a particular   
temperature `$T$'. The detailed calculations of $\langle\sigma v\rangle$ are discussed in \cite{clayton,fcz1,fcz2}. Any correction in the nuclear reaction rates 
on account of coulomb screening can change the nuclides abundances which directly affects the constraints on nucleosynthesis parameters. 

With this motivation, we first 
review the screening in the early universe using mean field theory and solve the relevant Poisson equation numerically in \sec{sieu}. This would give the reaction rate enhancement factor of two nuclei 
with charge $Z_1e$ and $Z_2e$ during nucleosynthesis. In \sec{prorates}, we determine the important reaction rates 
which play key role during the peak activity
of nucleosynthesis process in the formation and destruction of a particular nuclide. 
In \sec{esnp}, we further elaborate how coulomb screening 
affects nucleosynthesis parameters via nuclide abundances. This is qualitatively 
understood by restricting to the important reaction rates determined through processing rates. Finally, we also study
the impact on nuclide abundances and nucleosynthesis 
parameters under any deviation from mean field approach.  
      
\section{Coulomb screening in the early universe}
\la{sieu}
\begin{figure*}[t!]
\centering
\includegraphics[width=0.50\textwidth,scale=0.2]{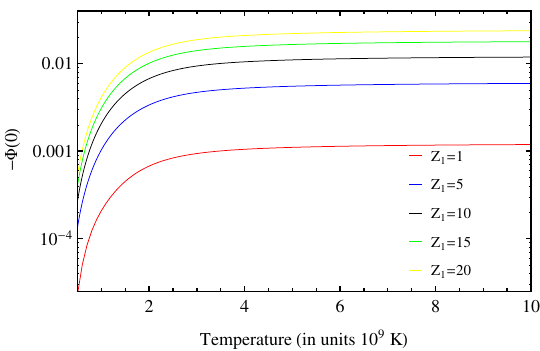}
\caption{Plot of -$\Phi(0)$ as a function of Temperature (in units 10$^9$K) for different
nuclei charges(Z$_1$).}
\label{y0vst9}
\end{figure*}
This section provides a brief review of coulomb screening which is 
required to determine the enhancement of the reaction rates between 
two interacting nuclei having charges $Z_1e$ and $Z_2e$ in the early universe. 
To determine the enhancement factor, we solve the Poisson equation numerically within the context of 
mean field theory on the similar lines as described in ref. \cite{salpeter} for the stellar cases. During nucleosynthesis,
when the temperature of the universe is of the order of few MeVs, the charged particle number density is dominated by the 
$e^\pm$ pairs. \\
This number density of electron-positron is given by the 
Fermi-Dirac distribution having momenta in the range $p$ to $p+dp$ at any temperature, $T$ with a 
chemical potential $\mu$, as \cite{weinberg}:
\bes
\label{I11}
n_{e^{\pm}}(p)=\frac{8\pi}{h^3}\int_{0}^{\infty}p^2\mathrm{d}p \left[1+\exp\left(\frac{E(p) \pm \mu}{kT}\right)\right]^{-1}
\ees
where $E(p)$=$p^2c^2+m_e^2c^4$ represents the energy of particles.\\ 
We replace variables, $p/m_ec$ = $\eta$, $m_ec^2/kT$=$z$ and rearrange \eqs{I11} as
\be
\label{noden2}
n_{e^{\pm}}(\mu,T)= \kappa I_{\pm}(\mu,T)
\ee
Here, $\kappa$ = $\pi^{-2}\left(m_ec/\hbar\right)^3$ and
\bes
\label{I1}
I_{\pm}(\mu,T)  = \int_{0}^{\infty}\frac{\eta^2\mathrm{d}\eta\ }{1+\exp\left(z(
\eta^2+1)^{1/2}\pm \mu/kT\right)}\nn
\ees
In this electron-positron plasma, we consider a nucleus 
having charge $Z_1e$ at the origin. We can then write the total 
electrostatic
potential, $U_{tot}(r)$   
at any point `r' from this origin.
This potential can be given by
the sum of coulomb potential of the bare nucleus and some unknown
potential $V(r)$:
\be
U_{tot}(r)= Z_1e/r + V(r)
\ee
Here $V(r)$ is the mean potential which is given by all 
the charged particles of plasma except the nucleus $Z_1$. This is also
termed as the screening potential because it screens the coloumb 
potential experienced by any interacting charged nuclei at that point. 
This reduction in coloumb potential through $V(r)$ would enhance the rate of interaction 
between $Z_1$ and some another interacting nucleus $Z_2$.  In ref. \cite{salpeter}, Salpeter
demonstrated that enhancement factor $E_s$ in the rate
of interaction between the nuclei $Z_1$ and
$Z_2$ will only require $V(0)$, i.e. the screening potential at the origin instead of $V(r)$.\\  
This enhancement factor is given by:
\be
\la{en11}
E_s = \exp\left({\frac{-Z_2eV(0)}{kT}}\right).
\ee
\eqs{en11} reduces our problem to solve screening potential at origin only.\\
\\
For the calculation of $V(0)$, we need to solve the Poisson equation 
which is given by
\be
\nabla^2U_{tot}(r) = -4\pi en(r) - 4\pi Z_1e\delta^3(r)
\ee 
This, then, reduces to the following form for the potential $V(r)$:
\be
\label{po1}
\nabla^2V(r) = -4\pi en(r).
\ee
Here, $n(r)$ represents the effective number density at distance `$r$' 
from the nucleus $Z_1$ due to the electrons and the positrons which is given by
\begin{align}
\label{rho1}
n(r) = n_{e^+}(\mu,T,U_{tot}(r)) - n_{e^-}(\mu,T,U_{tot}(r))
\end{align}
in terms of the modified Fermi-Dirac distribution. This modification occurs since the electrons and the positrons 
at `$r$'
experience an interaction energy equal to
$eU_{tot}(r)$.
Then their number densities will no longer be left uniform and will
arrange themselves in such a way 
that the total thermodynamic potential is uniform.
This modified distribution is given by:
\be
\label{noden2}
n_{e^{\pm}}(\mu,T,U_{tot}(r))=\kappa I_{\pm}(\mu,T,U_{tot}(r)) \nn
\ee
with
\be
\label{I1}
I_{\pm}(\mu,T,U_{tot})  =
\int_{0}^{\infty}\frac{\eta^2\mathrm{d}\eta\ }{1+\exp\left(z(
\eta^2+1)^{1/2}\pm (\mu/kT+eU_{tot}/kT)\right)}
\ee
In the early universe, the chemical potential corresponding to the electrons and positrons
is negligibly small, $\mu\approx0$ \cite{itoh,weinberg} and hence we can write,
%\begin{align}
%\label{noden1}
%n_{e^{\pm}}(\mu,T) &= \nn \\ 
%&c\int_{0}^{\infty}\frac{\eta^2\ 
%\mathrm{d}\eta\ }{1+\exp\left(z(
%\eta^2+1)^{1/2}\pm \mu/kT \right)}
%\end{align}
\be
\label{noden3}
n_{e^{\pm}}(0,T,U_{tot}(r)) =\kappa I_{\pm}(0,T,U_{tot}(r)).
\ee
We also need to impose the boundary conditions in order to solve for the potential. For this, we can take that at large distances,
\be
\label{vr}
U_{tot}(r)\rightarrow 0 ~~\mathrm{as}~ r\rightarrow\infty.
\ee
This makes sure that at sufficiently large distances, the number density of electrons and positrons returns to its field free value 
\be
\la{ninfty}
n_\infty = n_{e^-}(0,T,0)=n_{e^+}(0,T,0)=\kappa I(T).\\
\ee
with
\be
\la{it}
I(T)=\int_{0}^{\infty}\frac{\eta^2\ \mathrm{d}\eta\ }{1+\exp\left(z(
\eta^2+1)^{1/2} \right)}
\ee
Using Eqs. (\ref{rho1}) and (\ref{ninfty}) in \eqs{po1}, we have
\be
\label{vr}
\nabla^2V(r) = 4\pi en_\infty \left(\frac{I_{-}(0,T,U_{tot}(r))-I_{+}(0,T,U_{tot}(r))}{I(T)}\right)
\ee

Defining new variables,
\begin{align}
\label{all}
\Phi(r)\equiv\frac{eV(r)}{kT}, ~~R \equiv
&\left(\frac{kT}{4\pi e^2n_\infty}\right)^{1/2},\nn \\ 
x\equiv\frac{r}{R},~~  F \equiv &\frac{Z_1e^2}{RkT}, 
\end{align}
where, $R$ is defined as screening radius and
$F$ is the screening strength parameter which is defined by the ratio of 
coulomb interaction at screening radius and the average kinetic energy of the particle.
This parameter also
governs the weak ($F\ll$1), intermediate ($F\approx1$) and the strong screening ($F\gg1$) regimes. Using \eqs{all} in \eqs{vr}, we have
\begin{align}
\la{last1}
\nabla_x^2 \Phi(x) = & \frac{1}{I(T)}\left(\int_{0}^{\infty}\frac{\eta^2\ \mathrm{d}\eta\ }{1+\exp\left(z(
\eta^2+1)^{1/2} -F/x - \Phi(x)\right)}\right. \nn\\
&\left. -\int_{0}^{\infty}\frac{\eta^2\ \mathrm{d}\eta\ }{1+\exp\left(z(
\eta^2+1)^{1/2} +F/x + \Phi(x)\right)}\right)
\end{align}
Boundary conditions to solve \eqs{last1} are such that $\Phi(x)$ approaches a finite
value at the origin and approaches -$F/x$ as $x$ tends to infinity. To solve \eqs{last1} numerically it is found convenient to transform this equation to one involving $Z(x)$ = $x\Phi(x)$. Then resultant equation after this transformation becomes
\begin{table}\caption{Shows the numerical value of $\Phi_1(0)$ 
with Temperature (in units 10$^9$ K).}
\begin{center}
\begin{tabular}{|c|c|}\hline
~~~~Temperature~~~~ & ~~~~~-$\Phi_1(0)$~~~~~\\[0.5ex]
(in units 10$^9 K$) & ~~~(-$\Phi(0)$ for $Z_1$=1)~~~\\
\hline
$100.0$ & 1.24$\times10^{-3}$ 
\\ \hline
$50.0$ & 1.24$\times10^{-3}$  
\\ \hline
$20.0$ & 1.21$\times10^{-3}$ 
\\ \hline
$10.0$ & 1.19$\times10^{-3}$
\\ \hline
$5.0$ & 1.10$\times10^{-3}$
\\ \hline
$2.0$ & 6.74$\times10^{-4}$ 
\\ \hline
$1.0$ & 2.09$\times10^{-4}$
\\ \hline
$0.5$ & 2.33$\times10^{-5}$
\\ \hline
$0.3$ & 7.71$\times10^{-7}$
\\ \hline
$0.1$ & 4.51$\times10^{-15}$
\\ \hline
\end{tabular}\label{yvst9}
\end{center}
\end{table}
\begin{align}
\la{last}
\frac{d^2Z}{dx^2} =& \frac{x}{I(T)}\left(\int_{0}^{\infty}\frac{\eta^2 \mathrm{d}\eta}{1+\exp\left(z(
\eta^2+1)^{1/2} -(F+Z(x))/x\right)}\right.\nn\\
&\left.-\int_{0}^{\infty}\frac{\eta^2 \mathrm{d}\eta }{1+\exp\left(z(
\eta^2+1)^{1/2} +(F+Z(x))/x\right)}\right)
\end{align}
with the boundary conditions $Z(0)$ = 0 and $Z$ approaching -$F$ as $x$ tends to
infinity. We solve this equation numerically  on the lines of ref. \cite{salpeter} with these boundary conditions to
determine the value of $\Phi(0)$. We plot value of -$\Phi(0)$ as a function of temperature
for different values of $Z_1$ in \fig{y0vst9}.
This leads to an enhancement factor of reaction rates in the early universe
\be
E_s=\exp\left(-Z_2\Phi(0)\right)
\ee
Our numerical results are in good agreement with the results of Itoh et al \cite{itoh} and 
approximated by:
\be
E_s=\exp\left(-Z_1Z_2\Phi_{1}(0)\right)
\ee
where, $\Phi_1(0)$ is defined as $\Phi(0)$ for $Z_1=1$.
Numerical values of $\Phi_1(0)$ as function of temperature are shown in Table \ref{yvst9}.

\section{Processing Rates}
\label{prorates}

\begin{figure*}[p!]
\centering
\includegraphics[width=0.46\textwidth,scale=0.2]{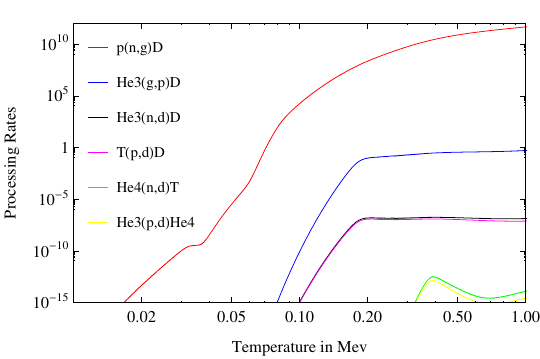}
\includegraphics[width=0.46\textwidth,scale=0.2]{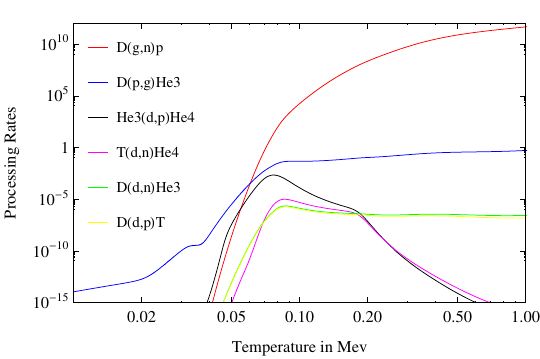}
\caption{Left Panel: Show the processing rates for Deuterium(D) production channels. \\Right Panel: 
 Show the processing rates for Deuterium(D) destruction channels.}
\label{dproc}
\end{figure*}
\begin{figure*}[p!]
\centering
\includegraphics[width=0.46\textwidth,scale=0.2]{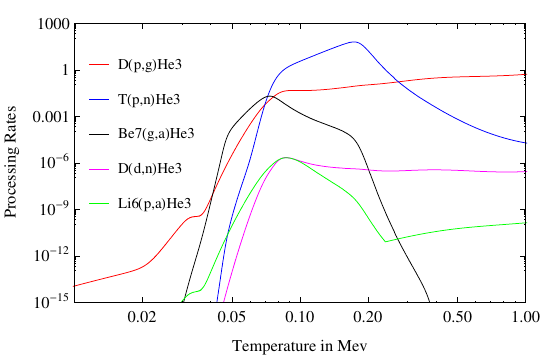}
\includegraphics[width=0.46\textwidth,scale=0.2]{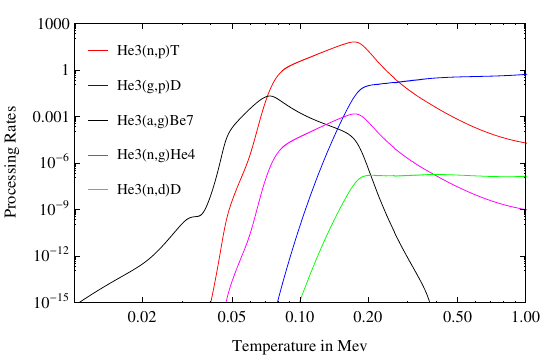}
\caption{Left Panel: Show the processing rates for Helium-3($^3$He) production channels. \\
Right Panel: 
Show the processing rates for Helium-3($^3$He) destruction channels.}
\label{he3proc}
\end{figure*}
\begin{figure*}[p!]
\centering
\includegraphics[width=0.47\textwidth,scale=0.2]{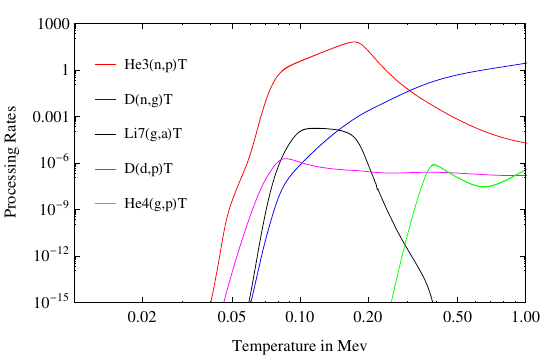}
\includegraphics[width=0.47\textwidth,scale=0.2]{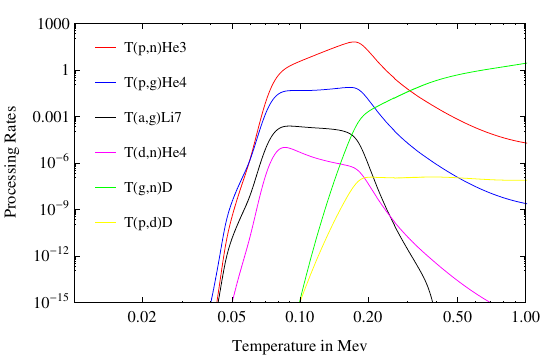}
\caption{Left Panel: Show the processing rates for Tritium(T) production channels.\\ Right Panel: 
 Show the processing rates for Tritium(T) destruction channels.}
\label{tproc}
\end{figure*}
\begin{figure*}[p!]
\centering
\includegraphics[width=0.46\textwidth,scale=0.2]{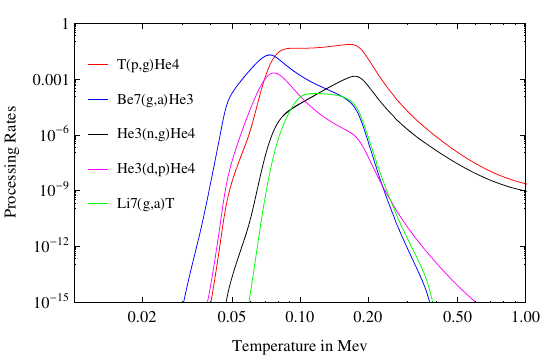}
\includegraphics[width=0.46\textwidth,scale=0.2]{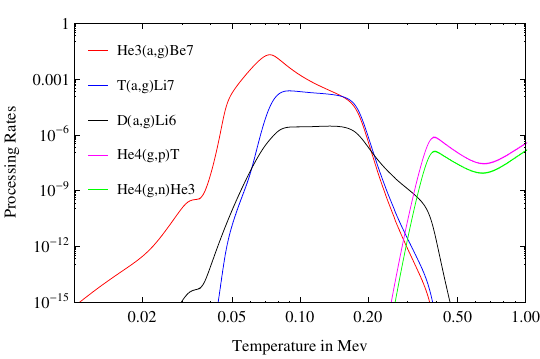}
\caption{Left Panel: Show the processing rates for Helium-4($^4$He) production channels.\\ Right Panel: 
 Show the processing rates for Helium-4($^4$He) destruction channels.}
\label{he4proc}
\end{figure*}
\begin{figure*}[p!]
\centering
\includegraphics[width=0.46\textwidth,scale=0.2]{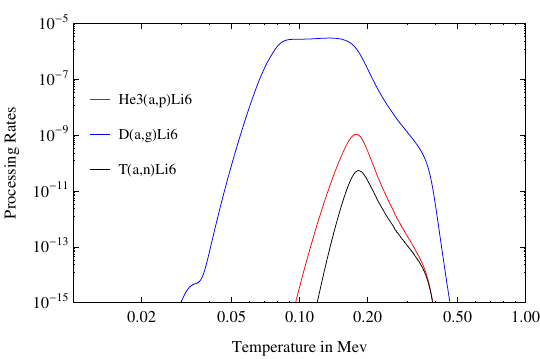}
\includegraphics[width=0.46\textwidth,scale=0.2]{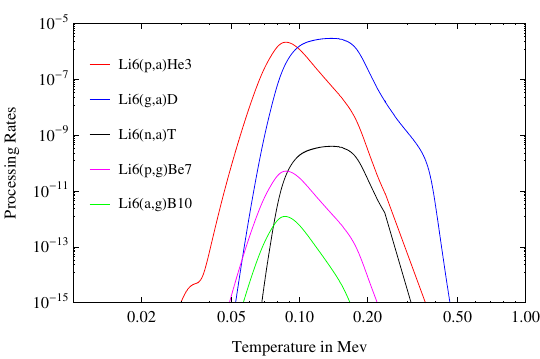}
\caption{Left Panel: Show the processing rates Lithium-6($^6$Li) production channels. \\Right Panel: 
 Show the processing rates for Lithium-6($^6$Li) destruction channels.}
\centering
\includegraphics[width=0.46\textwidth,scale=0.2]{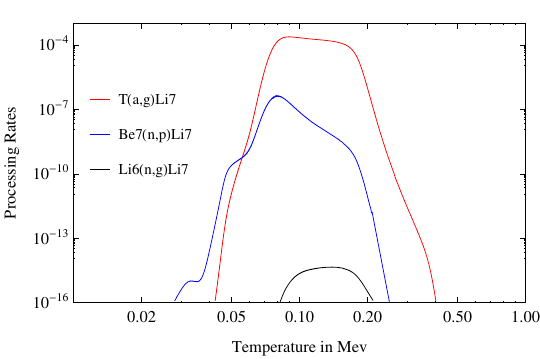}
\includegraphics[width=0.46\textwidth,scale=0.2]{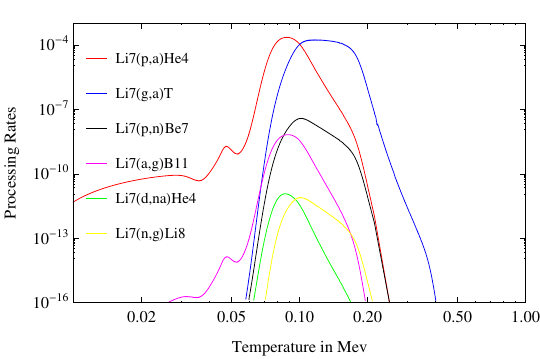}
\caption{Left Panel: Show the processing rates Lithium-7($^7$Li) production channels. \\Right Panel: 
 Show the processing rates for Lithium-7($^7$Li) destruction channels.}
\label{li7proc}
\end{figure*}
\begin{figure*}[t!]
\centering
\includegraphics[width=0.46\textwidth,scale=0.2]{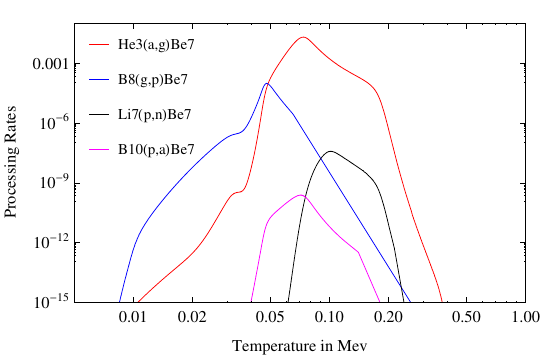}
\includegraphics[width=0.46\textwidth,scale=0.2]{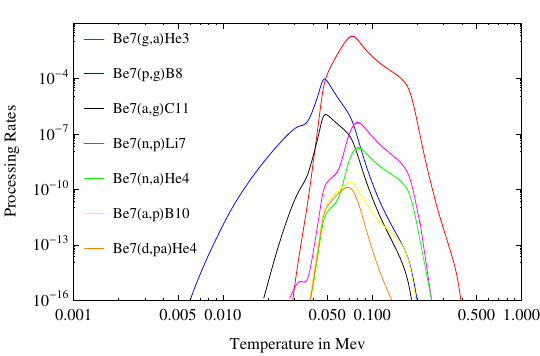}
\caption{Left Panel: Show the processing rates 
for Be-7($^7$Be) production channels.\\ Right Panel: 
Show the processing rates for Be-7($^7$Be) destruction channels.}
\label{be7proc}
\end{figure*}
To determine the primordial element abundances, the numerical calculation requires a network
of nuclear reaction rates. In LCN, this network determines 
130 element abundances via 557 reactions which is quite large in comparison with SBBN which has only 88 reactions.
But all these reactions are not equally important and only a
few of them dominate during nucleosynthesis.
These key reaction rates are important for
the qualitative understanding of screening effects during nucleosynthesis process 
and also help in the quantitative study of nucleosynthesis in such slow evolving models.
So it is important to discuss the method to determine 
the key reactions in the Linear coasting nucleosynthesis.\\ 
To know the key reaction rates, we have to calculate the processing rate 
for individual reaction as a function of temperature
and list them according to their importance during 
peak activity of nucleosynthesis. To figure out the processing rates,
we consider a reaction in which  
nuclei `$a$' is produced through `$c$' and `$d$' and destroyed on the interaction 
with nuclei `$b$'. This reaction can be written as:
\be
a+b\leftrightarrow c+d.
\ee  
We define the processing rate `P$_{\mathrm{prod}}(d(c,b)a)$' 
for the production of `$a$' via $c$ and $d$ as 
\begin{equation}
\mathrm{P}_{\mathrm{prod}}(d(c,b)a) \equiv \frac{\rho_bN_A\langle\sigma v\rangle_{cd}Y_cY_d}{H}
\end{equation}
where, $Y_i \equiv X_i/A_i$, $X_i$ is the mass fraction contained in the nuclei with atomic mass $A_i$, 
$\rho_b$ is the baryonic density of the universe and
$\langle\sigma v\rangle_{cd}$ represent the rate of this interaction.\\ 
Similarly, we can write the processing rate for the destruction of `$a$' 
nuclide on interacting with `$b$' nuclide as
\begin{equation}
\mathrm{P}_{\mathrm{dest}}(a(b,c)d) \equiv \frac{\rho_bN_A\langle\sigma v\rangle_{ab}Y_aY_b}{H}    
\end{equation}
where, $H$ is hubble expansion rate.\\
In Standard Big Bang Nucleosynthesis(SBBN), 
only 12 reactions play a significant role for the 
formation of lighter nuclei \cite{prates}. For the linear coasting nucleosynthesis, $^4$He and metallicity (Z) is produced upto observed level. 
The following reactions significantly affect the
production of $^4$He in LCN:
\begin{itemize}
\item D(p,g)$^3$He, $^3$He(d,p)$^4$He
\item T(p,n)$^3$He, T(p,g)$^4$He
\end{itemize} 
and $^4$He mainly destructs through the following reactions:
\begin{itemize} 
\item D(a,g)$^6$Li
\item T(a,g)$^7$Li
\item $^3$He(a,g)$^7$Be, $^3$He(a,p)$^6$Li
\end{itemize} 
Metallicity (Z) gets significantly produced by the 
following set of reactions:
\begin{itemize} 
\item $^7$Be(p,g)$^8$B, $^7$Be(a,g)$^{11}$C, $^7$Be(a,p)$^{10}$B   
\item $^7$Li(p,n)$^7$Be, $^7$Li(a,g)$^{11}$B
\item $^6$Li(p,g)$^7$Be, $^6$Li(a,g)$^{10}$B 
\end{itemize} 
as shown by Figs. \ref{dproc} - \ref{be7proc} where we have plotted the processing rates for the key elements as function of temperature in MeVs  . 
One of the important feature of these
plots is that the qualitative behaviour remains unchanged for 
different values of baryon to photon ratio($\eta_B$) and electron neutrino degeneracy parameter($\xi_e$). 

\section{Coulomb Screening and Nucleosynthesis Parameters}
\label{esnp}
Our previous work showed that the primordial 
element abundances are very sensitive to
two important nucleosynthesis parameters viz.; 
the baryon to photon ratio ($\eta_B$) and the electron neutrino degeneracy
parameter ($\xi_e$) \cite{parm2014}. 
The baryon to photon ratio ($\eta_B$) can also be written as:
\begin{align}
\eta_9=10^9\eta_B ~~;~~\eta_9=27.39\Omega_Bh^2
\end{align}
These nucleosynthesis parameters are tightly constrained
within the context of LCN (without screening) in ref.\cite{parm2014} to be
\begin{align}
\la{ppc}
\eta_9=3.927\pm0.292~~;~~\Omega_B=0.263&\pm0.026\nn\\
\xi_e=-2.165\pm0.171. 
\end{align}
These parameters are estimated such
that they produce the observationally inferred $^4$He abundance \cite{izotov1}
\be
\la{heab}
Y_p=0.254\pm 0.003 ~
\ee
and meet the minimum metallicity level (Z) required for the cooling and fragmentation process
of the observed low metal stars in the universe given as:
\be
\la{zab}
Z=Z_{cr}=10^{-6}Z_{\odot}.
\ee

\begin{figure*}[h!]
\centering
\includegraphics[width=0.46\textwidth,scale=0.2]{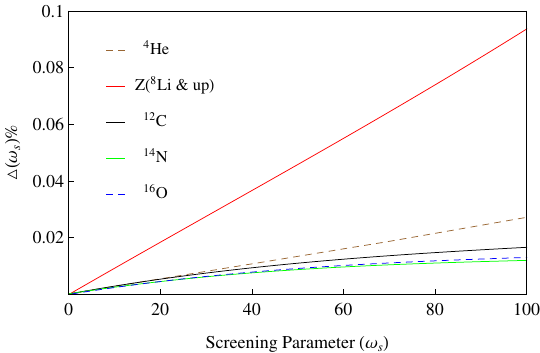}
\caption{Percentage variation ($\bigtriangleup(\omega_s)\%$) of predicted elemental abundances with the screening parameter($\omega_s$).}
\label{dev}
\end{figure*}

Here, $Z_\odot$ represents solar metallicity and the bounds for solar metallicity are given as \cite{chap07}: 
\be
0.0187~\le ~Z_\odot ~\le ~0.0239 .
\ee
With these constraints on parameters, one can also produce an appreciable amount of CNO in comparison 
with SBBN \cite{parm2014}. These levels of CNO essentially required to sustain CNO cycle in the massive stars \cite{cno}.
We now incorporate the screening effects discussed in \sec{sieu} enhancing all the nuclear reactions rates considered in the network of 557 reactions except the weak interactions, neutron and gamma induced reactions. 
However, it is found that this screening is unable to produce any measurable impact on primordial abundances of $^4$He and $Z$ as well as on nucleosynthesis parameter. 
This can qualitatively be apprehended by
using the processing rates from which
we get the reaction rates that dominantly affect $^4$He and $Z$ abundances as discussed in \sec{prorates}.
\\Most of the reactions in $^4$He production channel have $Z_1\times Z_2$=1
 with a maximum enhancement of:  
\be
\la{en1}
\exp(1.24\times10^{-3})\sim1.00124  
\ee
at high temperatures and it decreases sharply as the temperature decreases.
Similarly the metallicity production reaction rates are only enhanced
by:
\be 
\la{en2}
\exp(6\times1.24\times10^{-3})\sim 1.0075 
\ee
when produced via interaction between Lithium and $^4$He in which $Z_1\times Z_2$=6 and
production through Beryllium and $^4$He leads to an enhancement of  
\be
\la{en3}
\exp(8\times1.24\times10^{-3})\sim1.01 
\ee
where, $Z_1\times Z_2=8$. In this way, the coulomb screening can enhance
the reaction rates only upto one percent.
\\ Such an enhancement hardly impacts the previous 
results obtained without incorporating the screening. Thus, we can safely use the constraints on nucleosynthesis parameters as given in 
\eqs{ppc}.\\
\\However,
it is interesting to see 
how any deviation from the mean field approximation can affect 
the element abundances and the 
nucleosynthesis parameters.
To understand such deviations, we can artificially modify
our screening result by rescaling the screening potential by some screening parameter
$\omega_s$ which is defined in the following way:
\bes
E'_s = \exp \left( - \omega _s Z_2 Y(0) \right);~~
\omega_s=\ln{E_s}/\ln{E'_s}.
\ees   
with $\omega_s$ that can be considered to be lying in a range between 0 and 100.
We can then calculate percentage variation of any element abundance
as function of $\omega_s$ which is given by:
\be 
\bigtriangleup(\omega_s)=\frac{Y_A(\omega_s)-Y_A(\omega_s=0)}{Y_A(\omega_s=0)}.
\ee
where Y$_A(\omega_s)$ and Y$_A(\omega_s=0)$ represent any general 
element abundance for a screening parameter value $\omega_s$ and
without any coulomb screening respectively. 
We plot this percentage variation as a function of screening parameter
for the nuclides which are produced upto 
appreciable levels in such slowly evolving cosmologies as shown in \fig{dev}. This plot clearly shows that $^4$He is very less sensitive to the screening parameter in comparison
with metallicity ($Z$). Thus, as the screening parameter increases the dominant 
nuclear reaction rates involved in the production of metallicity are 
highly enhanced with respect to $^4$He as demonstrated in Eqs. (\ref{en1} - \ref{en3}). \\
\\
\begin{figure*}[h!]
\centering
\includegraphics[width=0.47\textwidth,scale=0.2]{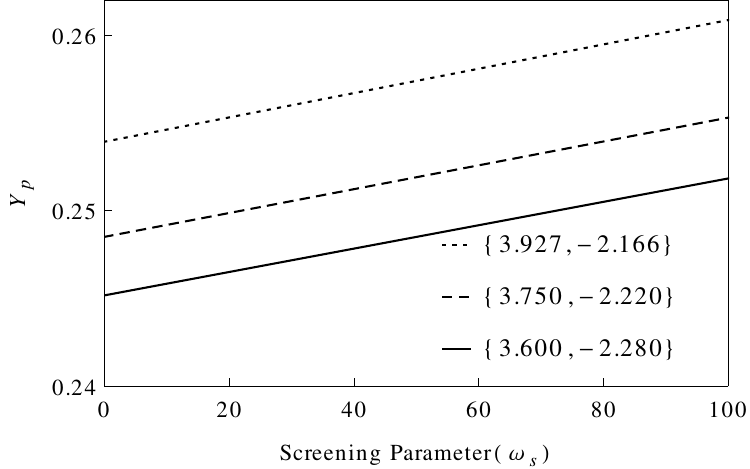}
\includegraphics[width=0.47\textwidth,scale=0.2]{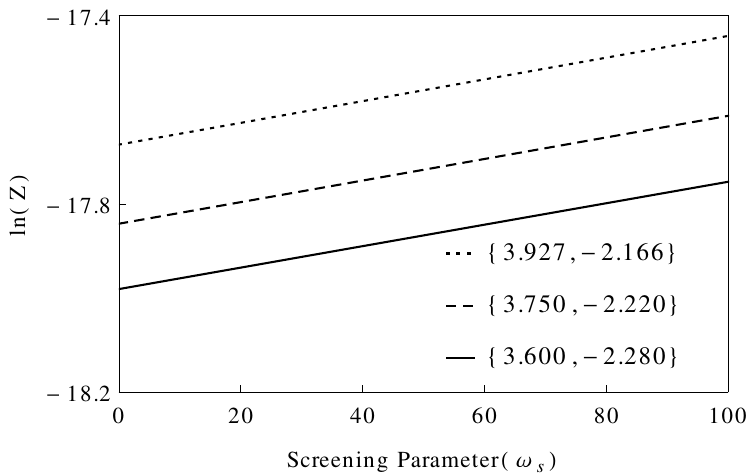}
\caption{Left and Right Panel demonstrate the linear response of $^4$He
abundance($Y_p$) and Metallicity $(\ln(Z))$ respectively with screening parameter($\omega_s$). 
For an individual linear plot we fix baryon to photon ratio, $\eta_9$ (in units 10$^9$)
and electron degeneracy parameter, $\xi_e$ and represented as $\{\eta_9,\xi_e\}$.}
\label{linear}
\end{figure*}
\begin{figure*}[h!]
\centering
\includegraphics[width=0.46\textwidth,scale=0.2]{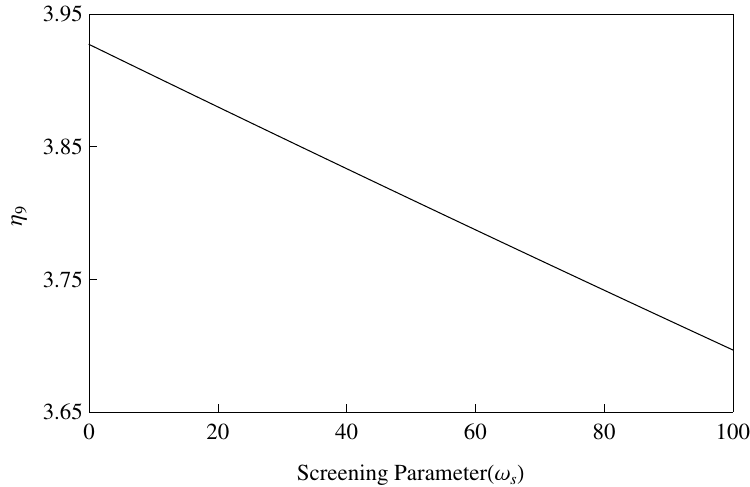}
\includegraphics[width=0.47\textwidth,scale=0.2]{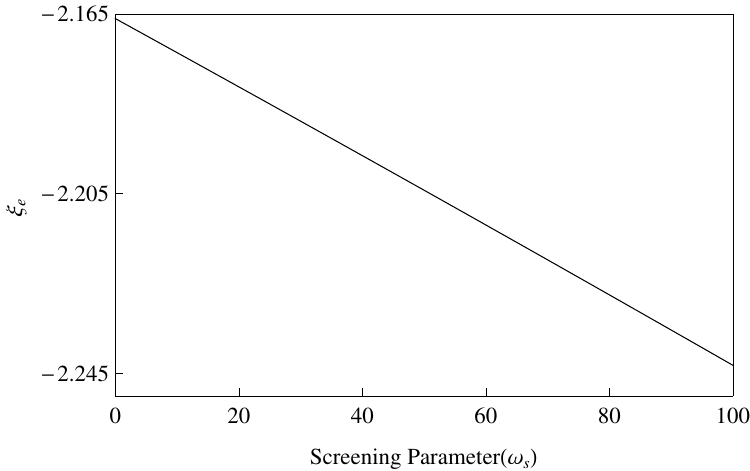}
\caption{Left Panel: Variation of baryon to photon ratio, $\eta_9$ (in units 10$^9$) as a function of screening parameter($\omega_s$).
Right Panel: Flow of electron neutrino degeneracy parameter($\xi_e$) with screening parameter($\omega_s$).}
\la{fop}
\end{figure*}
We also determine
the flow of nucleosynthesis parameters as a function of 
screening parameter by defining a general fitting function
for $^4$He($Y_p$) and $Z$ in which the screening parameter ($\omega_s$) is also included
\begin{align}
\la{hfitscr}
Y_p=A_0+B_0 \ln (\eta_9)+ C_0\xi_e +~~~~~~& \nn \\ \omega_s(A+B\ln(\eta_9)+C\xi_e)
\end {align}
\begin{align}
\la{zfitscr}
\ln(Z)=D_0+E_0 \ln (\eta_9)+ F_0\xi_e +~~~~~~& \nn\\ \omega_s(D+E\ln(\eta_9)+F\xi_e)
\end {align}
The coefficients are found to be:
\begin{align}
~~A_0=-0.5065;~B_0=0.3102;~C_0=-0.1554~~~~~~~~& \nn \\
A=3.516\times10^{-6};B=3.983\times10^{-5};C=-5.344\times10^{-6} & \nn \\
D_0=-30.4804;~E_0=6.2832;~F_0=-1.9555~~~~~~~~&\nn \\
D=1.902\times10^{-4};E=8.983\times10^{-4};F=-4.119\times10^{-4}\nn
\end{align}
In this model, we use the fact that the $Y_p$ and $\ln(Z)$ both are varying linearly with
$\omega_s$ as shown in \fig{linear}.
Using constraints of $Y_p$ and $Z$ values from Eqs.\ref{heab} and \ref{zab} in the above fitted equations we can 
solve for $\ln(\eta_9)$ and $\xi_e$ to get  
\begin{align}
\la{eta}
\ln(\eta_9)=\frac{1.368 - 7.232\times10^{-4}\omega_s - 3.138\times10^{-9}\omega_s^2}{1. - 8.783\times10^{-5}\omega_s - 3.138\times10^{-8}\omega_s^2}
\end{align}
\begin{align}
\la{xi}
\xi_e=-\left(\frac{2.165+6.414\times10^{-4}\omega_s+3.143\times10^{-8}\omega_s^2}{1. - 5.344\times10^{-5}\omega_s - 3.44\times 10^{-8} \omega_s^2}\right)
\end{align}
Note that, these solutions for the parameters are only valid for the range of $\omega_s$ lying between 0 to 100. We
plot $\eta_9$ and $\xi_e$ as a function of $\omega_s$  in \fig{fop}.
These plots abundantly show that the baryon to 
photon ratio ($\eta_9$) decreases and the electron neutrino degeneracy parameter($\xi_e$)
becomes more negative with $\omega_s$ respectively. 
This suggests that, even by enhancing the coulomb screening
effects 100 times, the nucleosynthesis parameters 
lie with in the error bars of the previous work \cite{parm2014}. Hence, we conclude that the
coulomb screening does not form any key ingredient
for the primordial nucleosynthesis even in slow evolving cosmological models.

\section{Discussion and Conclusion}

In this article, we have considered the effect 
of coulomb screening on the primordial nucleosynthesis in a linearly coasting universe via nuclear reaction rates. It was suggested in ref. \cite{pranavscr} that coulomb screening during the primordial nucleosynthesis may have an effect on the nuclear reaction rates in Linearly coasting universe. To determine the screening effect, we used the mean field approximation and solved the Poisson 
equation in the appropriate conditions that exist during primordial nucleosynthesis in linearly coasting universe. Due to lack of analytical solutions, we integrate this equation numerically using appropriate boundary conditions. These numerical results are in good agreement with the results of Itoh et al.

In order to interpret the whole effect of coulomb screening qualitatively
we also determined the key reaction rates from 557 reactions which significantly affect the element abundances. But it is important to note that the
screening is incorporated in all the reactions (excludes weak interactions as well as gamma and neutron induced reactions).  
Incorporating screening effects in 
LCN leads to negligibly small impact on nuclide abundances as well as on nucleosynthesis parameter \{$\eta_9,\xi_e$\}.  

We also show how any deviations from the mean field approximation can effect the nuclide abundances and parameters. 
This was studied by defining a screening parameter ($\omega_s$)
to boost the screening effects. The
 $^4$He abundance shows a very slight dependence on $\omega_s$
 in comparison with the metallicity ($Z$) levels. Even for large values of $\omega_s$, 
the nucleosynthesis
parameters cannot go beyond the error bounds of the previous work. This  
demonstrates that the coulomb screening does not play 
any key role in the primordial nucleosynthesis of slow evolving models.
In any case such large deviations 
are not justifiable on any physical grounds. 
All the previous works suggest that this screening parameter 
is in good agreement with $\omega_s \sim1$ and the claims in ref. \cite{pranavscr} are not compatible even with large values of the screening parameter.  This work strongly suggests that 
the observed amount of $^4$He and minimum metallicity requirements ($Z$)
can only be produced by a unique set of nucleosynthesis parameters in such a slow evolving cosmology. 

\bibliographystyle{utcaps}
\providecommand{\href}[2]{#2}\begingroup\raggedright\endgroup
\end{document}